\documentclass{article}
\usepackage{spconf,amsmath,graphicx}
\usepackage{iftex}
\ifXeTeX
  \usepackage{newtxmath}
  \usepackage[no-math]{fontspec}
\else
  \usepackage{amssymb}
\fi
\usepackage{booktabs}
\usepackage{needspace}
\usepackage{xcolor}
\usepackage[hyphens]{url}
\usepackage{orcidlink}
\hypersetup{hidelinks}

\graphicspath{{figures/}{generated/figures/}}

\input{generated/numbers.tex}

\makeatletter
\def\asom@stripci[#1]{#1}
\newcommand{\ci}[1]{\,(\expandafter\expandafter\expandafter\asom@stripci\csname res@#1\endcsname)}
\makeatother

\newcommand{\asom}{ASOM}

\title{Audio System One Model: Typed Probabilistic\\ Perception of Emotion in Speech}

\name{Jaime Bellver-Soler\textsuperscript{\orcidlink{0009-0006-7973-4913}}, Iv\'an Mart\'in-Fern\'andez\textsuperscript{\orcidlink{0009-0004-2769-9752}}, Anmol Guragain\textsuperscript{\orcidlink{0009-0009-8491-8663}}, Luis Fernando D'Haro\textsuperscript{\orcidlink{0000-0002-3411-7384}}}
\address{Speech Technology and Machine Learning Group (THAU)\\
Information Processing and Telecommunications Center, E.T.S.I. de Telecomunicación \\
Universidad Polit\'ecnica de Madrid, Spain}

\begin{document}
\ninept
\maketitle

\begin{abstract}
Speech applications ask many questions about the same voice: which emotion a listener perceives, how aroused the
speaker sounds, whether they sound frustrated. Classifiers answer fast, over one fixed label set; audio language
models (ALMs) take any question in text, but answer by generation, one pass per question, with poorly calibrated
probabilities. We present \asom{} (Audio System One Model), a System One model for speech: choice, score and yes/no
questions written at request time are answered in one pass, without generation, each with a probability
distribution. Frozen speech and text encoders feed a 6.2M-parameter reader with no per-class parameters, trained on listener
vote distributions. On MSP-Podcast, \asom{} matches a classifier on the same features and data, leads the best of
\res{alms.subset.count} zero-shot ALMs by \res{bestalm.subset.gap} and an ALM fine-tuned on the same data by
\res{omni3b_ftfull.subset.gap}, and answers 20 questions in \res{lat.asom_bkt.20q}\,ms,
\res{latalm.ratiomin}--\res{latalm.ratiomax} times faster than the ALMs. It also answers reworded, reduced and
new label sets, scales and gender, and on \res{zs.corpora} unseen corpora in \res{zs.langs} languages it leads the
best zero-shot ALM by \res{zs.bestalm.ctrl.gapabs} where the words carry no emotion; where they do, the ALMs mostly
read.
\end{abstract}


\section{Introduction}
\label{sec:intro}

\begin{figure}[t]
  \centering
  \includegraphics[width=0.88\columnwidth]{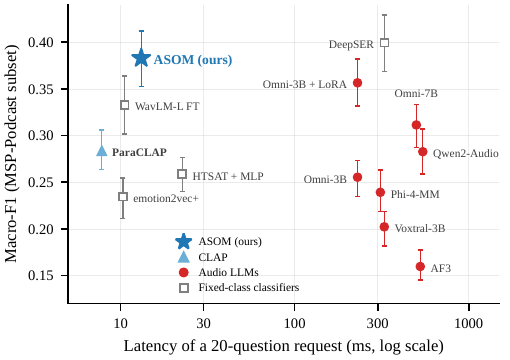}
  \caption{Accuracy vs.\ cost of 20 questions (clips of Table~\ref{tab:main}, 95\% intervals). Blue: one
  pass scores each option's text against the audio (\asom{} with cached criteria, CLAP). Red: audio LLMs, one LM
  pass per question, audio prefix shared. Grey: fixed-class classifiers.}
  \label{fig:positioning}
\end{figure}

Voice agents, call-centre analytics and content moderation ask many questions about the same clip, and a deployed
system usually fixes its questions, or a decision tree over them, in advance. Two families of models serve them, at opposite ends of a trade-off
(Fig.~\ref{fig:positioning}). Speech classifiers \cite{chen2022wavlm} are fast and calibrated, but a new emotion
word, a coarser label set or a new attribute needs a new model. Audio language models (ALMs)
\cite{chu2024qwen2audio,xu2025qwen25omni} take any question in text, but answer by generation: each question costs a
prefill, the implied probability is poorly calibrated, and they trail supervised models on emotion~\cite{zhang2026voxemo}.

\begin{figure*}[t]
  \centering
  \includegraphics[width=\textwidth,height=5cm]{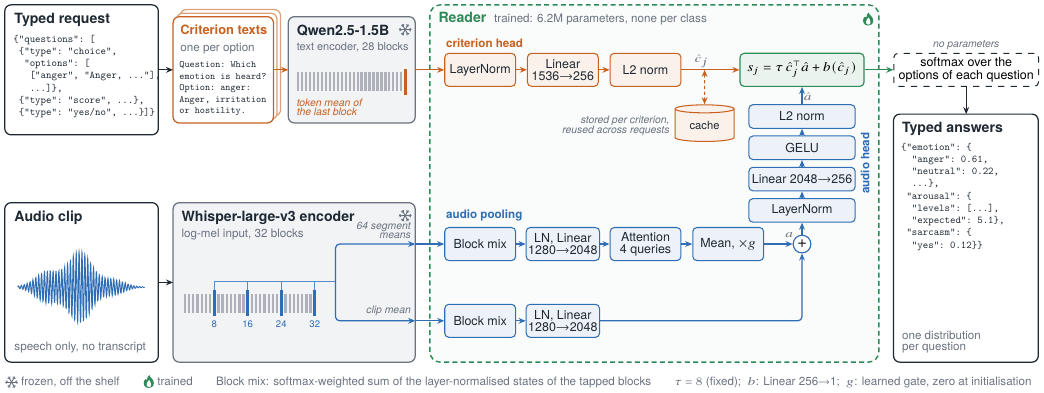}
  \caption{\asom{}. Each option becomes a criterion text, encoded by a frozen Qwen2.5-1.5B and cached, since it never
  sees the audio. The clip is encoded by a frozen Whisper-large-v3 and pooled from four blocks. The reader projects both
  to unit vectors, scores all criteria against the audio in one call (Eq.~\ref{eq:reader}), and a softmax per question
  gives the probabilities.}
  \label{fig:arch}
\end{figure*}

A third kind of model sits between the two: describe each label in words and score the input against that
description, as zero-shot classifiers of text, images and audio have long done (Sec.~\ref{sec:related}). TypeSafe
recently named such models \emph{System One} models \cite{almeida2026systemone}, after Kahneman's fast, intuitive
System~1 \cite{kahneman2011thinking}: a request carries typed questions (a \emph{choice} over described options, an
ordinal \emph{score}, a \emph{yes/no} proposition), each answered by a calibrated distribution in one pass, at the
latency of a small classifier. Their model, Jev, reads text only.

We present \asom{}, a System One model for speech: given a clip and questions written in text, it returns one
probability distribution per question. It is not a classifier with a fixed label set, since questions and options are written at request
time, nor an ALM, since it is not an autoregressive generative model, nor a contrastive model: each option becomes a text, a \emph{criterion},
scored against one audio vector, and the model is trained to reproduce the listeners' distribution of answers rather
than to rank captions. Because the labels are text, the same model can be asked a reduced or reworded label set, new
emotion words, an emotion scale or another attribute such as gender, without retraining; this instance is trained on
listener judgements of affect. The cost does not depend on the number of questions, and the criteria, independent of
the audio, are cached: a deployment with fixed questions pays only for the speech encoder. Our contributions:
\begin{itemize}
  \setlength\itemsep{0pt}
  \item \asom{}\footnote{Code and model weights will be publicly released upon acceptance. 
  This work has been submitted to the ICASSP for possible publication.}
  (Sec.~\ref{sec:method}): a 6.2M-parameter reader with no per-class parameters over frozen speech and text encoders,
  trained on listener vote distributions.
  \item A comparison on MSP-Podcast (Sec.~\ref{sec:res-msp}) with classifiers trained on the same data, six zero-shot
  ALMs, a fine-tuned ALM and three CLAP models, on the same clips.
  \item Tests of what a classifier cannot do (Secs.~\ref{sec:res-flex}--\ref{sec:res-zs}): reworded, reduced and new
  label sets, scales, gender and \res{zs.corpora} unseen corpora, with a transcript-only control separating
  hearing from reading.
\end{itemize}

\section{Related Work}
\label{sec:related}

\textbf{Speech emotion classifiers} put a fixed-class head on a pretrained speech encoder
\cite{chen2022wavlm,ma2024emotion2vecplus}, the strongest adding text streams over an ASR transcript
\cite{chatzichristodoulou2025medusa}. Training on listener vote distributions instead of majority labels
\cite{ando2018soft,chou2022typed} captures how listeners disagree; \asom{} keeps this supervision, but each class is a
text, not an output unit.

\textbf{Contrastive audio--text models} (CLAP) score a clip against any text \cite{wu2023clap}: MS-CLAP covers speech
emotion among 16 zero-shot tasks \cite{elizalde2024msclap}, paralinguistic models train on emotion captions
\cite{jing2024paraclap} or smoothed targets \cite{jing2026smoothclap}, and WavLink scores question--option texts
against a Whisper token \cite{kumar2026wavlink}. \asom{} shares their dual-encoder shape, but they return similarities
learned from captions or consensus labels, not calibrated answers learned from listener votes.

\textbf{Audio LLMs} take any question, but trail specialised models on emotion \cite{zhang2026voxemo}; fine-tuning
them towards annotator distributions \cite{yu2026disentangling} keeps a fixed label set and a billion-parameter
decoder.

\textbf{Labels as text and System One models.} Scoring written labels without generation is much older than its
current name: dataless classification embedded label names \cite{chang2008dataless}, entailment classifiers test each
label as a yes/no hypothesis \cite{yin2019entailment}, encoders score all the written options of a question in one pass
\cite{yang2022unimc} or several label sets at once \cite{zaratiana2025gliner2}, and dual encoders carried the idea to images
\cite{radford2021clip} and to audio, speech emotion included \cite{elizalde2023clap,xu2022zeroshot}. TypeSafe named
the class \emph{System One} models and popularised a typed request with Jev \cite{almeida2026systemone}, a text-only
model with no paper. Two concurrent audio prototypes \cite{balbay2026prosodia,barney2026circuit} use neither listener votes
nor shared benchmarks.

\section{Audio System One Model}
\label{sec:method}

Fig.~\ref{fig:arch} shows \asom{}: two frozen encoders turn the audio and the text of each option into vectors, and a
small trained \emph{reader} compares them and returns one probability distribution per question.

\textbf{Requests and criteria.} A request is one clip and a list of typed questions in JSON format
\cite{almeida2026systemone}: a \emph{choice} picks one of several options (e.g., which emotion is heard), a \emph{score} picks a level of an ordered scale (e.g., arousal from 1 to 7), and a \emph{yes/no} question asks whether a statement holds (e.g., is the speaker female?). Each option is written as one text, a \emph{criterion}: the question followed by the option's name and a short description (Fig.~\ref{fig:arch}). A choice question gives one criterion per option, a score one per level, and a yes/no question two.

\textbf{Encoders.} A frozen Qwen2.5-1.5B \cite{qwen25} processes each criterion: the layer-normalised mean of its
output token states is the criterion vector $c_j$, which does not depend on the audio and is cached. The frozen Whisper-large-v3 \cite{radford2023whisper} encodes the clip and from four of its 32 transformer blocks (8, 16, 24 and 32) we take the mean over the clip and the means of 64 time segments. A learned softmax mixes the four blocks, four learned queries attend over the segments, and their output is added to the clip mean through a zero-initialised gate, giving the audio vector $a$. If the request includes a transcript, a second frozen text encoder, RoBERTa-large, encodes it into one more input next to $a$.

\textbf{Reader.} Two small heads map both vectors to $\mathbb{R}^{256}$ and scale them to unit length:
$\hat c_j = W_c c_j/\lVert W_c c_j\rVert$ and $\hat a = h/\lVert h\rVert$, with $h=\mathrm{GELU}(W_a\,\mathrm{LN}(a))$.
Because both have unit norm, their dot product already is their cosine similarity, with no division by the norms,
and option $j$ gets
\begin{equation}
  s_j = \tau\,\hat c_j^{\top}\hat a + b(\hat c_j), \qquad p(j) = e^{s_j}\big/\textstyle\sum_{k\in Q} e^{s_k},
  \label{eq:reader}
\end{equation}
where $\tau=8$ is fixed, $b$, a linear function of the criterion alone, learns how often each option is chosen in
training, and the softmax runs over the options $Q$ of the question. A score also returns its expected level, and a
yes/no question the probability of its affirmative criterion. A request's criteria are all scored in one matrix product (63 criteria for 20 questions in 0.6\,ms), so the cost barely depends on how many questions; removing options renormalises the rest.

\textbf{Training.} Each question's target is the distribution of listeners' answers \cite{ando2018soft}:
the share picking each option, a score's rating histogram, or the share answering yes. The loss is the 
cross-entropy to this distribution (the log score, a strictly proper scoring rule), so the reader learns how listeners
disagree, not only their majority.

With few listeners per clip, these targets are noisy, so the reader also learns from a smoother one, built by
cross-fitting: the training clips are split into disjoint folds, auxiliary readers are trained on all folds but one,
and each predicts only the clips of the fold it never saw. Averaged per clip, these predictions form an ensemble
estimate of the listeners' answers; ensembles are better calibrated than single models
\cite{lakshminarayanan2017ensembles}, and distilling one into a single model retains much of that gain at the cost of one
model \cite{hinton2015distilling}. Because no auxiliary reader saw the clip it predicts, the target does not simply
copy that clip's votes. The loss adds the cross-entropy to this target to
the one to the votes; the final model is still one reader, and without this step its log loss
worsens (Sec.~\ref{sec:res-abl}). Only the reader is trained; the encoders remain frozen.

\section{Experimental Setup}
\label{sec:setup}

\begin{table}[t]
  \centering
  \caption{MSP-Podcast 2.0 Dev-select, the \res{n.subset} clips every system answers: 8-class macro-F1 (95\%
  interval), log loss against the votes (LL$_8$) and latency of 20 questions in ms (Sec.~\ref{sec:setup}). Omni: Thinker
  parameters. + transcript: an earlier reader on the fixed 30\,s window. Bold: best per column.}
  \label{tab:main}
  \scriptsize
  \setlength{\tabcolsep}{3pt}
  \begin{tabular}{llccc}
    \toprule
    System & Params & F1 (95\% CI) & LL$_8$ & ms \\
    \midrule
    \addlinespace[1.5pt]
    \multicolumn{5}{l}{\emph{Fixed-class classifiers}} \\
    \addlinespace[1pt]
    WavLM-L fine-tuned & 0.3B & \res{wavlm_ft.subset.f1}\ci{wavlm_ft.subset.f1ci} & \res{wavlm_ft.subset.ll} & \res{lat.wavlm_ft.20q} \\
    HTSAT (CLAP) + MLP & 0.2B & \res{clap_mlp.subset.f1}\ci{clap_mlp.subset.f1ci} & \res{clap_mlp.subset.ll} & \res{lat.clap_mlp.20q} \\
    emotion2vec+ large & 0.3B & \res{e2v.subset.f1}\ci{e2v.subset.f1ci} & \res{e2v.subset.ll} & \res{lat.e2v.20q} \\
    DeepSER & 3.3B & \res{deepser.subset.f1}\ci{deepser.subset.f1ci} & \res{deepser.subset.ll} & \res{lat.deepser.20q} \\
    \cmidrule(lr){1-5}
    \multicolumn{5}{l}{\emph{Audio LLMs: zero-shot, and fine-tuned with LoRA}} \\
    \addlinespace[1pt]
    Qwen2.5-Omni-3B & 4.7B & \res{omni3b.subset.f1}\ci{omni3b.subset.f1ci} & \res{omni3b.subset.ll} & \res{lat.omni3b.20q} \\
    Qwen2.5-Omni-7B & 8.9B & \res{omni7b.subset.f1}\ci{omni7b.subset.f1ci} & \res{omni7b.subset.ll} & \res{lat.omni7b.20q} \\
    Qwen2-Audio-7B & 8.4B & \res{qwen2audio7b.subset.f1}\ci{qwen2audio7b.subset.f1ci} & \res{qwen2audio7b.subset.ll} & \res{lat.qwen2audio7b.20q} \\
    Audio Flamingo 3 & 8.3B & \res{af3.subset.f1}\ci{af3.subset.f1ci} & \res{af3.subset.ll} & \res{lat.af3.20q} \\
    Phi-4-multimodal & 5.6B & \res{phi4mm.subset.f1}\ci{phi4mm.subset.f1ci} & \res{phi4mm.subset.ll} & \res{lat.phi4mm.20q} \\
    Voxtral-Mini-3B & 4.7B & \res{voxtral3b.subset.f1}\ci{voxtral3b.subset.f1ci} & \res{voxtral3b.subset.ll} & \res{lat.voxtral3b.20q} \\
    Omni-3B + LoRA (169k) & 4.7B & \res{omni3b_ftfull.subset.f1}\ci{omni3b_ftfull.subset.f1ci} & \res{omni3b_ftfull.subset.ll} & \res{lat.omni3b.20q} \\
    \cmidrule(lr){1-5}
    \multicolumn{5}{l}{\emph{One pass scoring each option: CLAP and ours}} \\
    \addlinespace[1pt]
    ParaCLAP & 0.3B & \res{clap_para.subset.f1}\ci{clap_para.subset.f1ci} & \res{clap_para.subset.ll} & \textbf{\res{lat.clap_para.20q}} \\
    \asom{} & 2.2B & \res{asom.subset.f1}\ci{asom.subset.f1ci} & \res{asom.subset.ll} & \res{lat.asom_bkt.20q} \\
    \asom{} + transcript & 2.6B & \textbf{\res{asom_text.subset.f1}}\ci{asom_text.subset.f1ci} & \textbf{\res{asom_text.subset.ll}} & \res{lat.asom_text.20q} \\
    \bottomrule
  \end{tabular}
\end{table}

\textbf{Training data.} Learning from votes needs corpora that release each listener's answer. Most emotion corpora
give one label per clip, usually the emotion an actor was asked to portray, which says nothing of how listeners
disagree; we keep them for testing. We train on three corpora that release individual votes, each with its own label set:
MSP-Podcast 2.0 \cite{busso2026msppodcast} (Train, 169k podcast clips with each listener's primary and secondary
emotions and 7-point arousal, valence and dominance), CREMA-D \cite{cao2014cremad}
(acted English, 6--12 listeners per clip) and the Italian Emozionalmente \cite{catania2025emozionalmente} (5 per clip).

\textbf{Training.} Each clip is asked four questions, with targets from the annotations: its \emph{primary
emotion}; \emph{arousal, valence, and dominance} (the
histogram of 7-point ratings); \emph{gender}; and \emph{yes/no} per emotion. Half of the steps paraphrase the questions and options, so the reader does not learn one wording. 
Cross-fitting (Sec.~\ref{sec:method}) uses 32 auxiliary readers (disjoint folds,
several seeds each); results average three seeds.

\textbf{Evaluation data.} The main benchmark is MSP-Podcast 2.0 Dev-select (\res{n.dev} clips, 553 speakers), a
speaker-disjoint development partition used for model selection. As the ALMs are slow, all systems answer the same speaker-stratified subset of \res{n.subset} clips (Table~\ref{tab:main};
on all \res{n.dev}, they score within the same intervals)
and a nested 1,000-clip subset (Table~\ref{tab:flex}). For transfer, we take up to 1,200 clips from each of
\res{zs.corpora} corpora unseen by \asom{} (\res{zs.clips} clips, \res{zs.langs} languages, acted and natural,
3--10 labels, 17 from EmoBox \cite{ma2024emobox}), each asked with its own labels; ten give speaker gender
(\res{n.genderref} clips). EmoNet-Voice Bench \cite{schuhmann2025emonet} (\res{nc.clips} synthetic clips,
\res{nc.emotions} emotions) and EARS \cite{richter2024ears} (\res{ears.clips} clips, \res{ears.emotions} emotions) add
emotions outside every training label set.

\textbf{Baselines.} Divided in three families:
\begin{itemize}
  \setlength\itemsep{0pt}
  \item \emph{Classifiers}, trained by us on MSP-Podcast: fine-tuned WavLM-large \cite{chen2022wavlm}, an MLP on
  CLAP's HTSAT audio encoder \cite{wu2023clap} and DeepSER, the four-stream block (WavLM, Whisper and two text encoders
  over its own transcript) of MEDUSA \cite{chatzichristodoulou2025medusa};
  plus emotion2vec+ large \cite{ma2024emotion2vecplus} off the shelf.
  
  \item \emph{Audio LLMs}, zero-shot: Qwen2.5-Omni 3B and 7B \cite{xu2025qwen25omni}, Qwen2-Audio-7B-Instruct
  \cite{chu2024qwen2audio}, Audio Flamingo 3 \cite{goel2025af3}, Phi-4-multimodal \cite{abouelenin2025phi4mm} and
  Voxtral-Mini-3B \cite{liu2025voxtral}. They get \asom{}'s question and option texts with lettered options, and we read
  the probabilities of the answer tokens instead of decoding. Omni-3B is also fine-tuned with LoRA (rank 16) on 20k and on all 169k Train clips with \asom{}'s targets.
  \item \emph{CLAP}: ParaCLAP \cite{jing2024paraclap}, MS-CLAP 2023 \cite{elizalde2024msclap} and LAION CLAP
  \cite{wu2023clap}.
\end{itemize}

\textbf{Metrics.} \emph{MSP-Podcast}: 8-class macro-F1 against the majority label; LL$_8$ against the votes; CCC for arousal, valence and dominance; ROC-AUC for yes/no. \emph{Other corpora}: macro-F1 on each corpus's labels, averaged over all unseen corpora (All) or those whose transcript is at chance (Voice); accuracy for gender. \emph{Cost}: latency of 20 questions on a 5.9\,s clip (warm median, one RTX 6000 Ada). Intervals: 95\% speaker-clustered bootstrap \cite{bisani2004bootstrap}; differences paired.

\section{Results}
\label{sec:results}

\subsection{Emotion recognition on MSP-Podcast}
\label{sec:res-msp}

Table~\ref{tab:main} compares all systems on the same clips. \asom{} (F1 \res{asom.subset.f1}) is above every
audio-only classifier; only DeepSER, which adds two text streams over its own transcript, is ahead (\res{deepser.subset.delta}); an earlier \asom{} reader given a transcript as an extra
input reaches \res{asom_text.subset.f1}. Every other open-label system trails \asom{}: the best zero-shot ALM,
\res{bestalm.subset.name}, by \res{bestalm.subset.gap}, ParaCLAP by \res{clap_para.subset.gap}, and Omni-3B fine-tuned
with LoRA on all 169k Train clips by \res{omni3b_ftfull.subset.gap}; the general-audio MS-CLAP and LAION CLAP are near
chance (\res{clap_ms.subset.f1} and \res{clap_laion.subset.f1}) and left out of Table~\ref{tab:main}. The probabilities differ even more: every
zero-shot ALM has a log loss above the class prior (\res{alms.subset.llmin} or more, against \res{prior.subset.ll}), so
a constant answer is closer to the listeners' votes, while \asom{} is well below it (\res{asom.subset.ll}).

Fig.~\ref{fig:positioning} adds the cost. \asom{} answers twenty questions in \res{lat.asom_bkt.20q}\,ms, as fast
as one, since all criteria are scored in one reader call; most of it is the Whisper
encoder (\res{lat.asom_bkt.whisper}\,ms). An ALM runs its language model for each question: even with the audio
prefix shared, twenty questions take \res{latalm.min20q}--\res{latalm.max20q}\,ms,
\res{latalm.ratiomin}--\res{latalm.ratiomax} times \asom{}. CLAP models are also flat and cheaper (ParaCLAP
\res{lat.clap_para.20q}\,ms), but less accurate. A new question also runs the text encoder once
(\res{lat.asom_bkt_encode.20q}\,ms for twenty, still below the fastest ALM); its criteria are then cached.

\subsection{Beyond a fixed label set}
\label{sec:res-flex}

A classifier answers only the label set it was trained on. \asom{} takes its labels as text, so the same model can be
asked what it was never trained to answer; Table~\ref{tab:flex} tests this against the systems that can do the same,
on the MSP-Podcast subset (left) and on corpora that none of them saw in training (right, Sec.~\ref{sec:res-zs}).
When every emotion is renamed with a synonym and every description is paraphrased (Syn.), \asom{}'s F1 is
\res{flex.asom.synpara.f1} (\res{flex.asom.synpara.delta} vs.\ Trained), still the best of the column, ahead of the
fine-tuned ALM (\res{flex.omni3b_ftfull.synpara.f1}). When the question itself is replaced by a new one (New\,q.),
its F1 is \res{flex.asom.instr2.f1} (\res{flex.asom.instr2.delta}), again the best of the column; no zero-shot ALM
exceeds \res{flexalm.native} in any column.

\begin{table}[t]
  \centering
  \caption{Systems that take their labels as text. Left, 1,000 MSP-Podcast clips: macro-F1 with the trained wording,
  with synonym names and paraphrased descriptions (Syn.) and with a reformulated question (New\,q.), each system on its best question (\asom{}: its decision question, as in
  Table~\ref{tab:main}); yes/no AUC for eight emotion words unseen in training (Words). Right: mean macro-F1 on the unseen corpora, all \res{zs.corpora}
  (All; chance \res{zs.chance.f1}) and the \res{zs.ctrl.n} whose transcript is at chance (Voice). Bold: best per
  column.}
  \label{tab:flex}
  \scriptsize
  \setlength{\tabcolsep}{2.2pt}
  \begin{tabular}{lcccccc}
    \toprule
    & \multicolumn{4}{c}{\emph{MSP subset}} & \multicolumn{2}{c}{\emph{Unseen corpora}} \\
    \cmidrule(lr){2-5}\cmidrule(lr){6-7}
    System & Trained & Syn. & New\,q. & Words & All & Voice \\
    \midrule
    Omni-3B & \res{flex.omni3b.native.f1} & \res{flex.omni3b.synpara.f1} & \res{flex.omni3b.instr2.f1} & \res{flex.omni3b.yesno_new_words} & \res{zs.omni3b.f1} & \res{zs.omni3b.f1ctrl} \\
    Omni-7B & \res{flex.omni7b.native.f1} & \res{flex.omni7b.synpara.f1} & \res{flex.omni7b.instr2.f1} & \res{flex.omni7b.yesno_new_words} & \res{zs.omni7b.f1} & \res{zs.omni7b.f1ctrl} \\
    Qwen2-Audio & \res{flex.qwen2audio7b.native.f1} & \res{flex.qwen2audio7b.synpara.f1} & \res{flex.qwen2audio7b.instr2.f1} & \res{flex.qwen2audio7b.yesno_new_words} & \res{zs.qwen2audio7b.f1} & \res{zs.qwen2audio7b.f1ctrl} \\
    AF3 & \res{flex.af3.native.f1} & \res{flex.af3.synpara.f1} & \res{flex.af3.instr2.f1} & \res{flex.af3.yesno_new_words} & \res{zs.af3.f1} & \res{zs.af3.f1ctrl} \\
    Phi-4-MM & \res{flex.phi4mm.native.f1} & \res{flex.phi4mm.synpara.f1} & \res{flex.phi4mm.instr2.f1} & \res{flex.phi4mm.yesno_new_words} & \res{zs.phi4mm.f1} & \res{zs.phi4mm.f1ctrl} \\
    Voxtral-3B & \res{flex.voxtral3b.native.f1} & \res{flex.voxtral3b.synpara.f1} & \res{flex.voxtral3b.instr2.f1} & \res{flex.voxtral3b.yesno_new_words} & \res{zs.voxtral3b.f1} & \res{zs.voxtral3b.f1ctrl} \\
    Omni-3B, LoRA & \res{flex.omni3b_ftfull.native.f1} & \res{flex.omni3b_ftfull.synpara.f1} & \res{flex.omni3b_ftfull.instr2.f1} & \textbf{\res{flex.omni3b_ftfull.yesno_new_words}} & \res{zs.omni3b_ftfull.f1} & \res{zs.omni3b_ftfull.f1ctrl} \\
    \midrule
    ParaCLAP & \res{flex.clap_para.native.f1} & \res{flex.clap_para.synpara.f1} & \res{flex.clap_para.instr2.f1} & \res{flex.clap_para.yesno_new_words} & \res{zs.clap_para.f1} & \res{zs.clap_para.f1ctrl} \\
    \asom{} & \textbf{\res{flex.asom.native.f1}} & \textbf{\res{flex.asom.synpara.f1}} & \textbf{\res{flex.asom.instr2.f1}} & \res{flex.asom.yesno_new_words} & \textbf{\res{zs.asom.f1}} & \textbf{\res{zs.asom.f1ctrl}} \\
    \bottomrule
  \end{tabular}
\end{table}

New emotions can also be added. Asked, as yes/no questions, about eight emotions that were never among its training
labels (Words), \asom{} reaches an AUC of \res{flex.asom.yesno_new_words} against the listeners' secondary votes,
close to the \res{flex.asom.yesno_native} it reaches on the trained emotions and on par with the best zero-shot ALM
(\res{flexalm.newwords}); only the fine-tuned ALM is higher (\res{flex.omni3b_ftfull.yesno_new_words}). A classifier
cannot be asked these questions at all. The same model also answers scales: asked for arousal, valence and dominance
on a 1--7 scale, it reaches a CCC of \res{flex.asom.ccc.arousal}, \res{flex.asom.ccc.valence} and
\res{flex.asom.ccc.dominance}, while the best of the other systems reaches \res{flex.omni3b_ftfull.ccc.arousal},
\res{flex.omni3b_ftfull.ccc.valence} and \res{flexalm.ccc.dominance}. Options can also be removed: since each is scored
on its own, \asom{} simply renormalises over the rest, which an ALM, reading all options together, does not
guarantee. Finally, gender, asked in a wording not seen in training, reaches an accuracy of \res{flex.asom.genderref} on ten corpora, below Omni-7B (\res{flex.omni7b.genderref}).

\subsection{Unseen corpora: hearing or reading}
\label{sec:res-zs}

On the \res{zs.corpora} corpora that \asom{} never saw (Table~\ref{tab:flex}, right), \asom{} reaches a mean
macro-F1 of \res{zs.asom.f1}, against \res{zs.bestalm.f1} for the best zero-shot ALM and \res{zs.omni3b_ftfull.f1}
for the fine-tuned one. Only emotion2vec+ is higher (\res{zs.e2v.f1}), but it was fine-tuned on EmoBox, which holds
17 of these corpora; on the other four the two are close.

Do the ALMs hear the emotion or read it? From the transcript alone, Omni-7B keeps 84\% of its F1 on MSP-Podcast
(\res{omni7b_text.subset.f1} of \res{omni7b.subset.f1}). On the \res{zs.lex.n} unseen corpora whose sentences reveal
the emotion, the best ALM leads \asom{} (\res{zs.bestalm.lex.f1} vs.\ \res{zs.asom.f1lex}); on the other
\res{zs.ctrl.n}, mostly acted with fixed sentences, the transcript is at chance and \asom{} is almost 50\% above the
best ALM (\res{zs.asom.f1ctrl} vs.\ \res{zs.bestalm.ctrl.f1}). The ALMs mostly read, as on emotionally incongruent
speech \cite{correa2026incongruent}; \asom{} hears. Emotions outside every training set tell the same story: on
EmoNet-Voice and EARS, whose scripts name the emotion, Omni-7B leads but keeps most of it from the transcript alone.
\asom{} ranks these labels sensibly (AUC \res{zs.asom.novelauc}) but seldom chooses them (F1 \res{zs.asom.novel}),
because its learned bias $b$ favours the trained labels.

\subsection{Ablations}
\label{sec:res-abl}

We reached this design by changing one part at a time, paired against the recipe on Whisper's fixed 30\,s window. \emph{Audio}:
Whisper-large-v3 beats WavLM-large and the audio tower of Qwen2.5-Omni (F1 \res{abl.wavlm.df1} and
\res{abl.omniaudio.df1}). Its deepest blocks matter (stopping at block 24 changes F1 by \res{abl.w24.df1}), and so do the
earlier ones (block 32 alone changes the unseen mean by \res{abl.w32.dzs}); without the time segments F1 changes
by \res{abl.segments.df1}, while one query instead of four changes the unseen mean by only \res{abl.q1.dzs}.
\emph{Window}: the shortest 10, 20 or 30\,s window instead of 30\,s cuts the 20-question latency from
\res{lat.asom_r1_local.20q} to \res{lat.asom_bkt.20q}\,ms at the same MSP-Podcast F1, changing the unseen mean by \res{abl.bkt.dzs};
no padding at all changes it by \res{abl.dyn.dzs}: short clips need Whisper's padding.
\emph{Text}: a class table in place of the criteria, one learned vector per class as in a classifier, on the same
features and data, keeps F1 (\res{whisper_mlp.full.f1} vs.\ \res{asom.full.f1}) and LL$_8$ (\res{whisper_mlp.full.ll}
vs.\ \res{asom.full.ll}) but lowers the unseen mean from \res{zs.asom.f1} to \res{zs.whisper_mlp.f1} (below \asom{} on
\res{zs.asom.winsctrl} of \res{zs.corpora} corpora): describing the classes in text costs no accuracy and helps
transfer. The descriptions matter more than the names: asked with the eight names alone, \asom{}'s F1 on the clips of
Table~\ref{tab:flex} falls by \res{flex.asom.bare_names.delta} (to \res{flex.asom.bare_names.f1}). The token mean beats the last token on held-out wordings (log loss \res{abl.lasttoken.dheld}). \emph{Training}: without distillation LL$_8$ worsens by
\res{abl.distill.dll}, without paraphrases the held-out-wording log loss by \res{abl.diversity.dheld}, and without
CREMA-D and Emozionalmente the unseen mean falls from \res{abl.datasetv1.zsctrl} to \res{abl.datasetv1.zs}.
\emph{Objective}: CLIP's contrastive loss on label captions instead of the votes changes F1
by \res{abl.clapobj.df1}, LL$_8$ by \res{abl.clapobj.dll} and yes/no AUC from \res{flex.asom_30s.yesno_native} to
\res{flex.asom_clip.yesno_native}, but raises the unseen mean by \res{abl.clapobj.dzs}: votes buy calibrated, typed
answers; a prior-free objective transfers better.

\section{Conclusion}
\label{sec:conclusion}
We asked whether a speech model can take its questions in text and still answer like a classifier: in one pass, with
calibrated probabilities. \asom{} shows that it can. Twenty typed questions cost little more than the speech encoder
(\res{lat.asom_bkt.20q}\,ms). On MSP-Podcast it
matches a classifier on the same features and data, so describing the classes in words costs no accuracy, and it is
more accurate, better calibrated and \res{latalm.ratiomin}--\res{latalm.ratiomax} times cheaper than six zero-shot
ALMs; an ALM fine-tuned on the same 169k clips still trails it in accuracy. What a classifier cannot do, \asom{} does
with the same weights: synonyms, paraphrases or a new question cost it under 0.04 of F1, eight emotion words unseen
in training reach an AUC of \res{flex.asom.yesno_new_words}, and on
\res{zs.corpora} unseen corpora in \res{zs.langs} languages it is the best open-label system, almost 50\% above the
best zero-shot ALM where the words carry no emotion.

The limits are as clear. These are closed tests on affect, and an ALM can be asked much that we did not test: who is
speaking, what the question refers to, what else is in the recording. \asom{} answers only what its data covers, three
corpora with listener votes, all about affect: labels outside them are ranked well (AUC \res{zs.asom.novelauc}) but
seldom chosen, gender trails Omni-7B, and where the words carry the emotion a transcript still matters. Data is the
bottleneck, not the interface, since nothing in the reader is tied to emotion. Our next steps are more data and a
larger model: multi-rater corpora in more languages and attributes, and an architecture scaled to any typed question
about a recording, to meet the ALMs one to one on every category, not only emotion.

\section{Compliance with Ethical Standards}
\label{sec:ethics}
We use public corpora under their licenses and collect no new data. The estimates describe how listeners judge a
voice, not the speaker's internal state, and should not drive high-stakes decisions.

\section{Acknowledgments}
\label{sec:ack}
This work is supported by the Program ``ACTIVIDADES I+D PROCESOS HUMANOS Y SOCIALES'' of the Comunidad de Madrid,
Spain, under Project PHS2024/PH-HUM-52 (Innovatrad-CM), by the Spanish Ministry of Science and Innovation through the project TRUSTBOOST (PID2023-150584OB-C21), funded by MCIN/AEI/10.13039/501100011033 and by the European Union ``NextGenerationEU/PRTR'', and by the European Innovation Council through Project EVIDENCE
(EIC No.~101192038). Claude (Anthropic) and GPT (OpenAI) assisted with code, literature search and writing; the
authors take full responsibility for the content.

\bibliographystyle{IEEEbib}
\bibliography{refs}

\end{document}